\documentclass[12pt]{article}
\begin{document}

\title{\bf Two-Loop Computation in Superstring
Theory\footnote{Based on the talk given at the International
Conference on String Theory in Beijing, August 17--19, 2002 and
the talk given at a Symposium on the occasion at the Founding of
the Interdisciplinary Center for Theoretical Study, University of
Science and Technology of China, Hefei, November 23--26, 2002.
This work is supported in part by fund from the National Natural
Science Foundation of China with grant Number 90103004. }}

\author{Chuan-Jie Zhu\footnote{e-mail: zhucj@itp.ac.cn} \\
Institute of Theoretical Physics\\
Chinese Academy of Sciences\\
P. O. Box 2735\\
Beijing 100080, P. R. China}

\maketitle

\begin{abstract}
In this paper I review some old and new works on the computation
of two-loop 4-particle amplitude in superstring theory. I also
present the proof  by Iengo, showing the vanishing of the term
related to the two-loop correction to the $R^4$ term. Finally I
will present some recent works on two-loop computation in
hyperelliptic language following the new gauging fixing method of
D'Hoker and Phong.
\end{abstract}


\section{String perturbation theory}

In the Polyakov approach to string perturbation theory, the
particle amplitude is given as
\begin{eqnarray}
A_n(k_i,\epsilon_i) & = & \sum_{\rm topologies} \int { {\cal
D}({\rm geometry }) {\cal D}({\rm string~coordinates}) \over {\rm
Vol.(symmetry~group)} } \nonumber \\
& & \times \int_\Sigma\prod_{i=1}^n {\rm d}^2 z_i
V(k_i,\epsilon_i;z_i,{\bar z}_i)\,{\rm e}^{-S},
\end{eqnarray}
where $V(k_i,\epsilon_i;z_i,{\bar z}_i)$ is the vertex for the
emission of the $i$-th particle with momentum $k_i$ and
polarization tensor $\epsilon_i$.

For bosonic string theory\footnote{We consider only oriented
closed string theory in this paper.} the action $S$ is given as
follows:
\begin{equation}
S = \int {\rm d}^2 \sigma \sqrt{g}g^{ab}\partial_a X\cdot
\partial_b X,
\end{equation}
where $X$ are string coordinates describing the embedding of
string in space-time, $ g_{ab}$ ($g^{ab}= (g^{-1})_{ab}$ is its
inverse) is the world-sheet metric and $\sigma^a$ ($a=1,2$) are
local coordinates on the world-sheet. It is not difficult to see
that this action has the following symmetries:
\begin{itemize}
\item   Reparametrization of the coordinates $\sigma^a$;
\item   Weyl rescaling of the metric $g_{ab}\to {\rm
e}^{\phi(\sigma)} g_{ab}$;
\item   Global symmetries of $X$: $X\to \Lambda \cdot X + A$.
\end{itemize}

The first step of string perturbation theory is to factor out all
the symmetries of the path integral and write the amplitude as an
integral over the moduli space.  To achieve this we decompose an
arbitrary variation of the metric $g_{ab}$ as follows:
\begin{equation}
\delta g_{\bar z\bar z} =\bigtriangledown_{\bar z} \bar v_{\bar z}
+ 2 g_{z\bar z}\delta y^i \mu_{\bar z}^{iz}, \label{eqzz}
\end{equation}
where $\mu_{\bar z}^{iz}$'s are called Beltrami differentials.
They can be contracted with the holomorphic 2-differentials
$\phi^i$ as follows:
\begin{equation}
\langle\phi^i,\mu^j\rangle =  \langle\mu^j,\phi^i\rangle =\int{\rm
d}^2 z \, \phi^i_{zz} \mu_{\bar z}^{jz} .
\end{equation}
The presence of the last term in (\ref{eqzz}) indicates that there
are non-trivial moduli which are denoted by $y^{i}$'s. It can be
proved that the moduli space can be given a complex structure and
so we will use $y^{i}$ as complex coordinates of the moduli space.
By factoring out the symmetry group the partition function at
genus $g\ge 2$ is given as follows\footnote{ $\langle\phi^i |
\phi^j\rangle $ is defined as follows:
\begin{equation}
\langle\phi^i | \phi^j\rangle = \int {\rm d}^2 z g^{z\bar z}
(\phi^i)^*\phi^j .
\end{equation} }:
\begin{equation}
Z_g = \int_{M_g}\prod_{i=1}^{3g-3} {\rm d}^2 y^i \, {|{\rm
det}\langle\phi^i, \mu^j\rangle |^2\over{\rm det}\langle\phi^i |
\phi^j\rangle } {\rm
det}'(\bigtriangledown_z\bigtriangledown^z_{-1})\int {\cal D}X
{\rm e}^{-S[X]} .
\end{equation}
Here   ${\rm det}'$ denotes an determinant omitting the zero
eigenvalue. By introducing ghost fields we have
\begin{equation}
Z_g = \int_{M_g}\prod_{i=1}^{3g-3} {\rm d}^2 y^i \,\int {\cal
D}[Xbc\bar b\bar c] \prod_{i=1}^{3g-3} |\langle b, \mu^i\rangle
|^2 \,
 {\rm e}^{-S[X]- S_{\rm gh}[b,c,\bar b, \bar c]} .
\end{equation}

The above can also be extended to superstring theory. Here the
natural formalism is a super extension of the world-sheet, i.e.
$N=1$ supergeometry. The analog of Riemann surfaces and moduli
space will be super Riemann surfaces and supermoduli space. A
complete account of this formalism can be found in
\cite{DHokerPhong1}. The measure was first derived in
\cite{Verlinde}. Here we only briefly recall the result.

For superstring theory the action is given as follows:
\begin{eqnarray}
S & = & {1 \over 4 \pi} \int _\Sigma {\rm d}^2z \sqrt{g}  \biggl (
{1\over 2} g^{mn} \partial_m x^\mu  \partial_n x^\mu + \psi ^\mu
\gamma ^m \partial_m
\psi ^\mu \nonumber \\
& & \hskip .8in - \psi ^\mu \gamma ^n \gamma ^m \chi _n \partial_m
x^\mu - {1 \over 4} \psi ^\mu \gamma ^n  \gamma ^m \chi_n (\chi_m
\psi ^\mu) \biggr  ).
\end{eqnarray}
This action is constructed so as to be invariant under
diffeomorphisms, local $N=1$ supersymmetry, Weyl and super Weyl
transformations of the worldsheet. In super conformal gauge it
reduces to the following form:
\begin{eqnarray}
S & = & {1\over 4\pi} \int {\rm d}^2 z ( \partial_z X\cdot
\partial_{\bar z} X - \psi_{+}\cdot \partial_{\bar z} \psi_{+\mu}
- \psi_-\cdot \partial_z \psi_- \nonumber \\
& & + {\chi_{\bar z}}^+\psi_+\cdot\partial_z X +
 {\chi_{  z}}^-\psi_-\cdot\partial_{\bar z} X - {1\over 2}
 {\chi_{\bar z}}^+{\chi_{  z}}^-\psi_+\cdot \psi_{-} ). \label{eqconf}
\end{eqnarray}
Following the
same strategy as in bosonic string theory,  we decompose an
arbitrary variation of ${\chi_{\bar z}}^+$ as follows:
\begin{equation}
{\chi_{\bar z}}^+ = \partial_{\bar z}\delta \epsilon(z,\bar z) +
\sum_{a=1}^{2g-2} \rho^a {\chi_{a\bar z}}^+, \label{eqrho}
\end{equation}
where $\rho^a$'s are odd super-moduli and ${\chi_{a\bar z}}^+$ are
a basis of super Beltrami differentials. The presence of the last
term (containing $\rho^a$) in (\ref{eqrho}) signifies the presence
of non-trivial super-moduli, i.e. we cannot set ${\chi_{\bar
z}}^+$ to zero completely by supersymmetry transformations. For
our convenience we can choose delta function for these
differentials:
\begin{eqnarray}
{\chi_{  z}}^- & = & \partial_z \bar\epsilon(z,\bar z)
 + \sum_a\bar\rho^a\delta^2(z-r_a),
 \\
{\chi_{\bar z}}^+ & = & \partial_{\bar z} \epsilon(z,\bar z) +
\sum_{a} \rho^a \delta^2(z-l_a).
\end{eqnarray}
It is important to choose $r_a\neq l_b$ for any $a,~b$. Otherwise
the last term in eq. (\ref{eqconf}) will give a non-vanishing
contribution, violating the splitting of the integrand into a
product of left part and right part.

The final formula for the measure after integration over
super-moduli $\rho^a$ is given as follows:
\begin{eqnarray}
Z_{g \ge 2 } & = & \int_{{\cal M}_g} \prod_i {\rm d}^2 m_i
\int {\cal D}[X\psi bc\beta
\gamma\bar b\bar c\bar\beta\bar\gamma] \, {\rm e}^{ -(S[X,\psi] +
S_{\rm gh}[b,c,\beta,\gamma] + S_{\rm gh}[\bar b,\bar c,
\bar\beta,\bar\gamma])}
\nonumber \\
& & \times \prod_a\delta( \beta(l_a)) \,J(l_a) \prod_i
\langle\mu_i,b\rangle \times (\hbox{ anti-holomorphic~part}) ,
\end{eqnarray}
where $J(z)$ is the total supercurrent:
\begin{equation}
J(z) = - { 1\over 2} \psi(z)\cdot \partial_zX(z) +{1\over2}b(z)
\gamma(z) -{3\over 2}\beta(z)\partial_zc(z) -
\partial_z\beta(z)c(z).
\end{equation}
Here the subscript ``$-$" for $\psi$ is omitted as the variable
$z$ will indicated weather it is left-handed or right-handed.

\section{One-loop amplitudes}
Green and Schwarz computed one-loop amplitudes before 1982. You
can find these computations in their papers
\cite{GreenSchwarz1,GreenSchwarz2} and the book of Green, Schwarz
and Witten \cite{GreenSchwarzWitten}. A modern version can be
found also in the big book of Polchinski \cite{Polchinski}. What
we will do here is to repeat this calculation in hyperelliptic
language of the Riemann surface \cite{Dixon, Zamolodchikov, Radul,
Knizhnik2, Gava}.

First we remind that a genus-g Riemann surface, which is the
appropriate world sheet for one and two loops, can be described in
full generality by means of the hyperelliptic formalism. This is
based on a representation of the surface as two sheet covering of
the complex plane described by the equation:
\begin{equation}
y^2(z) = \prod_{i=1}^{2g+2} ( z- a_i), \label{covering}
\end{equation}
The complex numbers $a_{i}$, $(i=1,\cdots,2g+2)$ are the $2g+2$
branch points, by going around them one passes from one sheet to
the other. For one-loop ($g=1$) one of them represent the moduli
of the genus 1 Riemann surface (the torus) over which the
integration is performed, while the other three can be arbitrarily
fixed.

Also in $g=1$, by choosing a canonical homology basis of cycles we
have the following list of three even spin structures:
\begin{eqnarray}
& & s_2 \sim \left[\begin{array}{c} 1 \\ 0 \end{array} \right]
\sim
(a_1 a_2|a_3 a_4), \\
& &  s_3 \sim \left[\begin{array}{c} 0 \\
0 \end{array}\right] \sim (a_1 a_3|a_2 a_4), \\
& &  s_4 \sim \left[\begin{array}{c} 0 \\ 1 \end{array}\right]
\sim (a_1 a_4|a_2 a_3) .
\end{eqnarray}
We will denote an even spin structure as $(A_1  A_2|B_1 B_2)$. The
one odd spin structure $s_1$ gives a vanishing contribution to the
$n$-particle amplitude for $n\le 4$ because of the presence of
zero modes.

For even spin structures there is no odd
supermoduli and the $n$-particle amplitude (for $n\le4$) is given as
follows:
\begin{equation}
A_n(k_i,\epsilon_i)   =   \int {\rm d} \mu \prod_{i=1}^n {\rm d}^2
z_i \sum_{s}\eta_s Q_s \langle \prod_{i=1}^n
V(k_i,\epsilon_i;z_i,\bar z_i) \rangle_s \times ({\rm right~part})
\label{eqoneloop},
\end{equation}
where the measure ${\rm d}\mu$ is (in terms of the branch points):
\begin{equation}
{\rm d}\mu = { 1\over T^6} \, {\prod_{i=1}^4 {\rm d}^2 a_i/{\rm
d}V_{pr} \over |\prod_{i < j} a_{ij}|^4 } , \label{eq20}
\end{equation}
for type II superstring theory. $Q_s = (A_1-A_2)(B_1-B_2)$ are
spin structure dependent factors from the determinants and
$\eta_s$ are phases which guarantee modular invariance. Here in
hyperelliptic Riemann surface language modular invariance is just
the invariance under the permutations of the 4 branch points. The
other quantities appearing in eq. (\ref{eq20})  are $T = \int {
{\rm d}^2 z \over |y(z)|^2 }$ and ${\rm d}\mu = {{\rm d}^2 a_i
{\rm d}^2 a_j {\rm d}^2 a_k \over |a_{ij}a_{jk}a_{ki}|^2}$. In
(\ref{eqoneloop}) the summation over spin structures was written
only for the left part. Appropriate factor of $Q_{s'}$ and the
summation over $s'$ are vaguely denoted as (right part).

By explicit calculation we have the following identities:
\begin{eqnarray}
& & \Lambda =  \sum_s \eta_s Q_s = 0 , \label{eqeta} \\
& & \sum_s \eta_s Q_s \langle \psi\psi(z_1) \,
\psi\psi(z_2)\,\psi\psi(z_3)  = 0 , \\
& & \sum_s \eta_s Q_s \langle \psi\psi(z_1) \,
\psi\psi(z_2)\,\psi\psi(z_3)\,\psi\psi(z_4)   \ne 0 , \\
& & \sum_s \eta_s Q_s S_s(z_1,z_2)  S_s(z_2,z_3) S_s(z_3,z_4)
S_s(z_4,z_1) = {\prod_{i < j}(a_i-a_j) \over \prod_{i=1}^4 y(z_i)
} . \label{eqlast}
\end{eqnarray}
Here in eq. (\ref{eqeta}) we can arbitrarily fix $\eta_2 = 1$ and
the other two $\eta_s$'s can be deduced by the requirement of
modular invariance. In fact  modular invariance here means the
antisymmetric property of the cosmological constant $\Lambda$
under the interchange of two arbitrary branch points $a_i
\leftrightarrow a_j$.

By using these identities we found that the cosmological constant
and $n$-point function ($n\le3$) are zero point-wise in moduli
space. By using eq. ({\ref{eqlast}) we can compute the left
(holomorphic) part of the 4-point amplitude and the result is:
\begin{equation}
({\rm 4-point:~left~part}) = ({\rm kinematic~factor}) \times
 {\prod_{i < j}(a_i-a_j) \over \prod_{i=1}^4 y(z_i)
} ,
\end{equation}
and the full amplitude (for type II superstring theory and bosons
from the NS-NS sector) is then
\begin{eqnarray}
A_4(k_i,\epsilon_i) & = &  K(k_i,\epsilon_i) \int { 1\over T^6} \,
{\prod_{i=1}^4 {\rm d}^2 a_i/{\rm d}V_{pr} \over |\prod_{i < j}
a_{ij}|^2 } \nonumber \\
& & \quad \times \prod_{i=1}^4 { {\rm d}^2 z_i \over |y(z_i)|^2 }
\, {\rm exp}[ \sum_{i <j} k_i\cdot k_j \langle X(z_i,\bar z_i)
X(z_j,\bar z_j)\rangle ],
\end{eqnarray}
where $K$ is the same kinematic factor as found in tree-level
amplitude. This result is identical with the one obtained in the
$\Theta$-function formalism
\cite{GreenSchwarz2,GreenSchwarzWitten}:
\begin{eqnarray}
A_4(k_i,\epsilon_i) & = & K(k_i,\epsilon_i) \int_F { {\rm d}^2
\tau \over ({\rm Im}\tau)^6} \nonumber \\
& & \quad \times \prod_{i=1}^4 { {\rm d}^2 z_i' } \, {\rm exp}[
\sum_{i <j} k_i\cdot k_j \langle X(z_i',\bar z_i') X(z_j',\bar
z_j')\rangle ].
\end{eqnarray}
This can be proved by using the following transformations:
\begin{eqnarray}
& & {\partial \tau \over \partial a_i} = {i\pi \over 2}
(\hat\omega(a_i))^2, \\
& & { {\rm d} z_i \over K y(z_i) } = {\rm d} z_i', \\
& & T = |K|^2 {\rm Im}\tau , \quad \omega(z) = { 1\over K y(z)}.
\end{eqnarray}
For $k \to 0 $ the 4-particle amplitude can be computed exactly. A
non-zero result is obtained apart from the kinematical factor:
\begin{equation}
A_4(k_i,\epsilon_i) \to  K(k_i,\epsilon_i) \times {\pi \over 3} \,
.
\end{equation}
This shows that the one-loop correction to the $R^4$ term
\cite{GrossWitten} is non-vanishing.

\section{Two-loop amplitudes}

Here at two loops we use the same strategy as at one loop. There
are ten even spin structures which we list in the following:
\begin{eqnarray}
s_1 \sim \left[ \begin{array}{cc} 1 & 1\\ 1 & 1 \end{array}
\right]  \sim (a_1 a_2 a_3|a_4 a_5 a_6), & & s_2 \sim \left[
\begin{array}{cc} 1 & 1\\ 0 & 0 \end{array} \right] \sim
(a_1 a_2 a_4|a_3 a_5 a_6), \nonumber
\\
s_3 \sim \left[ \begin{array}{cc} 1 & 0\\ 0 & 0 \end{array}
\right]  \sim (a_1 a_2 a_5|a_3 a_4 a_6), & & s_4 \sim \left[
\begin{array}{cc} 1 & 0\\ 0 & 1 \end{array} \right] \sim
(a_1 a_2 a_6|a_3 a_4 a_5), \nonumber
\\
s_5 \sim \left[ \begin{array}{cc} 0 & 1\\ 0 & 0 \end{array}
\right]  \sim (a_1 a_3 a_4|a_2 a_5 a_6), & & s_6 \sim \left[
\begin{array}{cc} 0 & 0\\ 0 & 0 \end{array} \right] \sim
(a_1 a_3 a_5|a_2 a_4 a_6), \nonumber
\\
s_7 \sim \left[ \begin{array}{cc} 0 & 0\\ 0 & 1 \end{array}
\right]  \sim (a_1 a_3 a_6|a_2 a_4 a_5), & & s_8 \sim \left[
\begin{array}{cc} 0 & 0\\ 1 & 1 \end{array} \right] \sim
(a_1 a_4 a_5|a_2 a_3 a_6), \nonumber
\\
s_9 \sim \left[ \begin{array}{cc} 0 & 0\\ 1 & 0 \end{array}
\right]  \sim (a_1 a_4 a_6|a_2 a_3 a_5), & & s_{10} \sim \left[
\begin{array}{cc} 0 & 1\\ 1 & 0 \end{array} \right] \sim
(a_1 a_5 a_6|a_2 a_3 a_4). \nonumber
\end{eqnarray}
We will denote an even spin structure as $(A_1 A_2 A_3|B_1 B_2
B_3)$. By convention $A_1= a_1$. As in one loop the spin structure
dependent factor from determinants are encoded in the following
factor \cite{GavaIengoSotkov}:
\begin{equation}
Q_s = \prod_{i <j} (A_i-A_j)(B_i-B_j),
\end{equation}
which is a degree 6 homogeneous polynomials in $a_i$.

At two loops there are two odd supermoduli and this gives two
insertions of supercurrent  at two different points $x_1$ and
$x_2$.  As it was done in \cite{GavaIengoSotkov} it is quite
convenient to choose these two insertion points as the two zeros
of a holomorphic abelian differential which are moduli independent
points on the Riemann surface. In hyperelliptic language these two
points are the same points on the upper and lower sheet of the
surface.

As in one loop we can prove the following identities:
\begin{eqnarray}
& & \Lambda = \sum_s \eta_s Q_s \sum_{i=1}^3 (A_i^n - B_i^n) = 0,
\\
& & \sum_s   \eta_s Q_s \left\{ {u(z_1) u(z_2)\over u(z_3)u(z_4) }
- {u(z_3) u(z_4)\over u(z_1)u(z_2) } \right\} = 0,
\\
& & \sum_s   \eta_s Q_s \left\{ {u(z_1) u(z_2)\over u(z_3)u(z_4) }
+ {u(z_3) u(z_4)\over u(z_1)u(z_2) } \right\}
\sum_{i=1}^3 (A_i^n - B_i^n) \nonumber \\
& &   =  {2 P(a) \prod_{i=1}^2 (z_i-z_3)(z_i-z_4)
\over \prod_{i=1}^4 y(z_i)} \left\{ \begin{array}{ll} 1  & n =1,\\
\sum_{i=1}^6 a_i - \sum_{k=1}^4 z_k & n=2. \end{array} \right.
\label{eqlasta}
\end{eqnarray}

By using these identities we found that the cosmological constant
and $n$-point function ($n\le3$) are zero point-wise in moduli
space \cite{IengoZhu1,Zhu}. Moreover for the 4-particle amplitude
the ``cross contraction'' terms \begin{equation} \langle
J(x_1)\psi(z_i) \rangle \langle J(x_2) \psi(z_j) (\cdots) \rangle,
\end{equation}
also gives vanishing contribution after summation over spin
structures. By using eq. ({\ref{eqlasta}) we can compute
explicitly the 4-particle amplitude. The difficult part is to
compute the ghost contributions and this has been achieved in
\cite{IengoZhu2,Zhu}. The full amplitude (for type II superstring
theory and bosons from the NS-NS sector) is \cite{IengoZhu2,Zhu}:
\begin{eqnarray}
& & AII(k_i,\epsilon_i) = c_{II} \, K\, \int { {\rm d}^2 a_1 {\rm
d}^2 a_2 {\rm d}^2 a_3 \, |a_{45}\,a_{46}\,a_{56}|^2 \over T^5
\prod_{i<j}^6 |a_{ij}|^2 } \prod_{l=1}^4 {{\rm d}^2 z_l
(r-z_l)(\bar{s}-\bar{z}_l) \over |y(z_l)|^2 }
\nonumber \\
& & \qquad \times \left\{ \left(I(r)\bar{I}(\bar{s}) + {5\over
4}\left( {\pi \over T \, y(r)\bar{y}(\bar{s}) } \int {{\rm d}^2 w
(r-w) (\bar{s}-\bar{w})\over |y(w)|^2 }\right)^2 \right) \langle
\prod e^{i k\cdot X}\rangle \right.
\nonumber \\
& & \qquad \qquad + {1\over 16}  \left\langle \partial X(r+)\cdot
\partial X(r-) \bar{\partial} X(\bar{s}+) \cdot \bar{\partial}
X(\bar{s}-)
 \prod e^{i k\cdot X}\right\rangle
\nonumber \\
& & \qquad \qquad \left. - {1\over 16} \left\langle \partial
X(r+)\cdot \partial X(r-) \bar{\partial} X(\bar{s}+) \cdot
\bar{\partial} X(\bar{s}-)\right\rangle \left\langle \prod e^{i
k\cdot X}\right\rangle \right\} , \label{EqAIIcomplete}
\end{eqnarray}
where $c_{II}$ is an overall constant which should be fixed by
unitarity and $K$ is the standard kinematic factor (for
$\epsilon^{\mu\nu}_i = \epsilon^{\mu}_i \tilde{\epsilon}^{\nu}_i$)
\cite{GreenSchwarz1, IengoZhu2, Zhu}:
\begin{eqnarray}
K & = & K_R \cdot K_L,
\\
K_R & = & -{1\over 4} (st\epsilon_1\cdot\epsilon_3
\epsilon_2\cdot\epsilon_4 +su \epsilon_2\cdot\epsilon_3
\epsilon_1\cdot\epsilon_4 + tu \epsilon_1\cdot\epsilon_2
\epsilon_3\cdot\epsilon_4 )
\nonumber \\
& & + {1\over 2} \, s \, ( \epsilon_1\cdot k_4 \epsilon_3\cdot k_2
\epsilon_2 \cdot \epsilon_4  + \epsilon_2\cdot k_3 \epsilon_4\cdot
k_1 \epsilon_1 \cdot \epsilon_3
\nonumber \\
& & \qquad + \epsilon_1\cdot k_3 \epsilon_4\cdot k_2 \epsilon_2
\cdot \epsilon_3  + \epsilon_2\cdot k_4 \epsilon_3\cdot k_1
\epsilon_1 \cdot \epsilon_4  )
\nonumber \\
& & + {1\over 2} \, t \, ( \epsilon_2\cdot k_1 \epsilon_4\cdot k_3
\epsilon_1 \cdot \epsilon_3  + \epsilon_3\cdot k_4 \epsilon_1\cdot
k_2 \epsilon_2 \cdot \epsilon_4
\nonumber \\
& & \qquad + \epsilon_2\cdot k_4 \epsilon_1\cdot k_3 \epsilon_3
\cdot \epsilon_4  + \epsilon_3\cdot k_1 \epsilon_4\cdot k_2
\epsilon_1 \cdot \epsilon_2  )
\nonumber \\
& & + {1\over 2} \, u \, ( \epsilon_1\cdot k_2 \epsilon_4\cdot k_3
\epsilon_2 \cdot \epsilon_3  + \epsilon_3\cdot k_4 \epsilon_2\cdot
k_1 \epsilon_1 \cdot \epsilon_4
\nonumber \\
& & \qquad + \epsilon_1\cdot k_4 \epsilon_2\cdot k_3 \epsilon_1
\cdot \epsilon_4  + \epsilon_3\cdot k_2 \epsilon_4\cdot k_1
\epsilon_1 \cdot \epsilon_2  ),
\\
K_L & = & K_R(\epsilon \to \tilde{\epsilon}).
\end{eqnarray}
Here $s$, $t$ and $u$ are the standard Mandelstam variables for
the 4-gravitons. Defining
$t_{\mu_1\nu_1\mu_2\nu_2\mu_3\nu_3\mu_4\nu_4}$ as follows:
\begin{equation}
K_R = t_{\mu_1\nu_1\mu_2\nu_2\mu_3\nu_3
 \mu_4\nu_4} \epsilon_1^{\mu_1} k_1^{\nu_1}
\epsilon_2^{\mu_2} k_2^{\nu_2}\epsilon_3^{\mu_3} k_3^{\nu_3}
\epsilon_4^{\mu_4} k_4^{\nu_4},
\end{equation}
the $R^4$ term is given as follows:
\begin{equation}
R^4 = t^{\mu_1\nu_1\mu_2\nu_2\mu_3\nu_3 \mu_4\nu_4}
t^{\rho_1\sigma_1\rho_2\sigma_2\rho_3\sigma_3\rho_4\sigma_4}
R_{\mu_1\nu_1\rho_1\sigma_1}R_{\mu_2\nu_2\rho_2\sigma_2}
R_{\mu_3\nu_3\rho_3\sigma_3}R_{\mu_4\nu_4\rho_4\sigma_4}.
\end{equation}
The other functions appearing in (\ref{EqAIIcomplete}) are
\begin{equation}
T (a_i, \bar{a}_i)  =  \int { {\rm d}^2 z_1 {\rm d}^2 z_2
 |z_1 - z_2|^2 \over |y(z_1)y(z_2)|^2 },
\label{Eqtwo}
\end{equation}
which is proportional to the determinant of the period matrix (see
ref. \cite{Zhu}), and
\begin{eqnarray}
I(r)& = & -{1\over 2} \sum_{i<j}^6 { 1\over r -a_i}
 {1\over r -a_j} -
{1\over 4} \sum_{i<j}^3 { 1\over r -a_i} {1\over r -a_j}
\nonumber \\
& & + {1\over 8} \left( \sum_{i=1}^6 {1 \over r - a_i} - 2
 \sum_{i=1}^3 {1 \over r - a_i}
\right) \sum_{l=1}^4 { 1\over r -z_l}
\nonumber \\
& & + {1\over 4}\sum_{i=1}^6 {1\over r -a_i} \sum_{i=1}^3 {1\over
r -a_i} - {5\over 4} \sum_{i=1}^6 {1\over r -a_i} {\partial \over
\partial a_i} \ln T, \label{Eqthree}
\end{eqnarray}
We will discuss the property of the above amplitude in the next
section.

Green and Gutperle \cite{GreenGutperle} have studied the effects
of $D$-instantons. They conjectured that there are perturbative
corrections to the $R^4$ term only from one-loop. Green and Sethi
\cite{GreenSethi} proved this conjecture from the SUSY constraints
on type IIB supergravity. In \cite{Green2},  Green,  Gutperle and
Vanhove argued the vanishing of the perturbative correction to the
$R^4$ term by using the one loop amplitude in eleven dimensions.
If our computation is correct we should expect a zero contribution
from two loops. We will discuss this issue in the next two
sections.

\section{Explicit modular invariant 4-particle amplitude}

In \cite{IengoZhu3}, we get an explicit modular invariant
4-particle amplitude\footnote{For simplicity we set $k \to 0$
everywhere except in the kinematic factor. The derivation also
goes with non zero $k$.}:
\begin{eqnarray}
& & AII_0   = c_{II}\,   K\, \int { \prod_{i=1}^6 {\rm d}^2 A_i
\over {T}^5 \prod_{i<j}^6 |A_{ij} |^2 } \prod_{l=1}^4 {{\rm d}^2
Z_l(r- Z_l) (\bar{s} - \bar{Z}_l) \over | Y(Z_l)|^2  }
\nonumber \\
& & \quad \times \prod_{i=1}^3\delta^2( Z_i- Z_i^0)
{\prod_{i<j=1}^3 |Z_i^0-Z_j^0|^2 }
\nonumber \\
& &\quad  \times  \left\{ I_M(r) \bar{I}_M(\bar{s})+ {5 \over 4}
\left( {\pi \over T \, Y(r)\bar{Y}(\bar{s} ) } \int {{\rm d}^2 v
(r-v) (\bar{s}-\bar{v})\over |Y(v)|^2 }\right)^2 \right\}.
\label{Eqlastz}
\end{eqnarray}
where $I_M(x)$ is given as follows:
\begin{eqnarray}
I_M(x) & = & {1\over 4} \sum_{i=1}^6 { 1\over (x- A_i)^2 }-
{1\over 4} \sum_{i<j}^6 { 1\over x -A_i}
 {1\over x -A_j} -   {1\over 8} \sum_{i=1}^6 {1 \over x- A_i}
 \sum_{l=1}^4 { 1\over x -Z_l}
\nonumber \\
& & + {1\over 4}\sum_{k<l=1}^4 {1\over x -Z_k} {1\over x -Z_l} -
{5\over 4} \sum_{i=1}^6 {1\over x-A_i} {\partial \over \partial
A_i} \ln T, \label{Eqthreem}
\end{eqnarray}
Again, $r$ and $\bar s$ are arbitrary and the result does not
depend on them.

By using the above formula, one can prove that the 4-particle
amplitude is finite. We refer the reader to \cite{IengoZhu3} for
details.

\section{The vanishing of the 4-particle amplitude in the limit
$k\to 0$}

In this section we present the proof of the vanishing of the
perturbative correction to the $R^4$ term at two loops by Iengo
\cite{Iengo}. The method used is a direct calculation of the
obtained integral in moduli space.

First we take the two insertion points to be $r=z_1^0$ and
$s=z_2^0$.\footnote{We change all $Z_i$ to $z_i$ and $A_i$ to
$a_i$.} Then we set $z_1^0 \to \infty$, $z_2^0 \to 0$ and $z_3^0 =
x$ and see how the integral varies with $x$. By making these
choices we have:
\begin{equation}
  AII_0 = -\bar x|x|^2 \int\ {\rm d}\mu\int {{\rm d}^2z~\bar z\over
|y(z)y(0)y(x)|^2}\, \left[ R\, L-{1\over \bar x}\, L-x\, R+{x\over
\bar x}\right] ,
\end{equation}
where
\begin{eqnarray}
{\rm d}\mu & \equiv & {\prod_{i=1}^6 {\rm d}^2a_i \over
T^5\prod_{i<j}^6 |a_i-a_j|^2},  \\
L & \equiv &  {1\over 2}\sum_{i=1}^6{a_i}-{z}, \qquad \quad R
\equiv {1\over 2}\sum_{i=1}^6{1\over\bar a_i}-{1\over\bar z}.
\end{eqnarray}
It can be checked that $AII_0$ is independent of $x$, by rescaling
the integration variables, and that the integral is convergent, by
the same analysis summarized in the Sect. 3 of ref.
\cite{IengoZhu3} (see the tables there).

We begin by observing that:
\begin{equation}
{1\over |y(0)|^2}\, \left[ {1\over 2} \, \sum_{i=1}^6{1\over\bar
a_i} \right]  =-\sum_{i=1}^6 {\partial \over \partial \bar
a_i}{1\over |y(0)|^2}.
\end{equation}
Therefore, the following identity holds for an integral expression
which we call $Q$:
\begin{eqnarray}
Q & \equiv &  \int {\rm d} \mu\int { {\rm d}^2z~\bar z\over
|y(z)y(0)y(x)|^2} \left[ {1\over 2}\sum_{i}{1\over\bar a_i}
\right]\, (L-x) \nonumber \\
& = & \int{\rm d} \mu\int {{\rm d}^2 z~\bar z\over |y(0)|^2}
\sum_i{\partial \over
\partial \bar a_i}{L-x\over |y(z)y(x)|^2}.
\end{eqnarray}
by integrating by parts and observing that $\sum_{i=1}^6 \left[
{\partial \over \partial \bar a_i}{1\over T^5\prod |a_i-a_j|^2}
\right]  =0$. \footnote{We note that
\begin{equation}
\int {\rm d}^2 a \, {\partial \over \partial \bar a}  f(a,\bar a)
\times { 1 \over a^n} = \int {\rm d}^2 a \, {\partial \over
\partial \bar a} \left[ f(a,\bar a) \times { 1 \over a^n} \right] ,
\end{equation}
may not exactly integrate to 0. One should carefully study the
singularity at $a=0$. }

Also,
\begin{equation}
\sum_{i=1}^6  {\partial \over \partial \bar a_i}\left[{L-x\over
|y(z)y(x)|^2} \right]  =-(L-x)\left[ {\partial \over \partial \bar
z}+{\partial \over \partial \bar x} \right] {1\over |y(z)y(x)|^2}.
\end{equation}
Thus, by integrating by parts in ${\rm d}^2 z$ we get:
\begin{equation}
Q=\int{\rm d}\mu\int {\rm d}^2z~{(L-x)\over |y(0)y(z)|^2} \left[
1-\bar z{\partial \over \partial \bar x} \right] \, {1\over
|y(x)|^2}.
\end{equation}
The result of the above steps is that:
\begin{eqnarray}
& & \hskip -2cm \int{\rm d}\mu\int {\rm d}^2z~{\bar z\over
|y(0)y(z)y(x)|^2} R(L-x)
\nonumber \\
& = & \int{\rm d}\mu\int {\rm d}^2z~{(L-x)\over |y(0)y(z)|^2}
\left[ 1-\bar z{\partial \over \partial \bar x}-{\bar z\over \bar
z} \right] {1\over |y(x)|^2}   \nonumber \\
& = & -{\partial \over \partial \bar x}\int{\rm d}\mu\int {\rm
d}^2 z~{\bar z\over |y(0)y(z)y(x)|^2} (L-x).
\end{eqnarray}
By using the previous results, we see that we can write our
amplitude $AII_0$ in the following form:
\begin{eqnarray}
AII_0 & =& \bar x |x|^2\left[  {\partial \over \partial \bar x}
 +{1\over \bar x} \right] \,
\int{\rm d}\mu\int { {\rm d}^2z~\bar z \, L\over |y(z)y(0)y(x)|^2}
\nonumber \\
&-&|x|^4\left[  {\partial \over \partial \bar x}
 +{1\over \bar x} \right] \,\int
{\rm d} \mu\int { {\rm d}^2z~\bar z\over |y(z)y(0)y(x)|^2}.
\label{finampl}
\end{eqnarray}
Now we perform a rescaling of the integration variables:
\begin{equation}
a_i \to x a_i, \qquad \qquad z \to x z,
\end{equation}
and we have
\begin{eqnarray}
\int{\rm d}\mu\int { {\rm d}^2z~\bar z\, L\over |y(z)y(0)y(x)|^2}
& = & { 1 \over |x|^2} \, \int{\rm d}\mu\int { {\rm d}^2z~\bar z\,
L\over |y(z)y(0)y(1)|^2}, \\
\int {\rm d} \mu\int { {\rm d}^2z~\bar z\over |y(z)y(0)y(x)|^2} &
= & {1\over x|x|^2} \,   \int {\rm d} \mu\int { {\rm d}^2z~\bar
z\over |y(z)y(0)y(1)|^2}.
\end{eqnarray}
Noting that
\begin{equation}
\left[   {\partial \over \partial \bar x} +{1\over \bar x} \right]
\, {1\over \bar x} = 0,
\end{equation}
we conclude that the amplitude $AII_0$ is zero, and therefore
there is no perturbative contribution to the invariant $R^4$ term
at two loops.

This is in agreement with the indirect argument of Green and
Gutperle \cite{GreenGutperle}, Green, Gutperle and Vanhove
\cite{Green2}, and Green and Sethi \cite{GreenSethi} that the
$R^4$ term does not receive perturbative contributions beyond one
loop. Recently, Stieberger and Taylor \cite{Stieberger}  also used
the result of \cite{IengoZhu2} to prove the vanishing of the
heterotic two-loop $F^4$ term. For some closely related works we
refer the reader to the reviews \cite{Green3, Kiritsis}.

\section{Recent works at two-loop superstring: work done with
Zhu-Jun Zheng and Jun-Bao Wu}

The problem with previous gauge fixing is the residual dependence
on the insertion points. This is due to the non-supersymmetric
projection of supermoduli to (even) moduli. In some recent papers
\cite{DHokerPhong2, DHokerPhong3, DHokerPhong4, DHokerPhong5},
D'Hoker and Phong  carried out a new supersymmetric gauge fixing
for the two-loop case. The even moduli are super period-matrix.
For a recent review see ref. \cite{DHokerPhong6}. We have done
explicit computations by using the newly obtained results for the
cosmological constant and $n$-particle amplitude for all $ n\le4$.
Here we will briefly present the main results and refer the reader
to \cite{AllZhu1, AllZhu2, AllZhu3} for full details. We note that
D'Hoker and Phong also do these computations by using
$\theta$-function formalism and the split gauge which is different
from our choice of the gauge \cite{DHokerPhong7, DHokerPhong8}.
Although the final results are exactly the expected, their
computation is quite difficult to follow because of the use of
theta functions.\footnote{In \cite{Lechtenfeld5}, the two-loop
4-particle amplitude was also computed by using theta functions.
Its relation with the previous explicit result \cite{IengoZhu2}
has not been clarified.}  Also modular invariance is absurd in
their computations because of the complicated dependence between
the 2 insertion points (the insertion points are also spin
structure dependent). Of course the final result should be
independent on the gauge choice.

\subsection{Some conventions and formulas}

We will follow the notations of D'Hoker and Phong
\cite{DHokerPhong2,DHokerPhong3,DHokerPhong4,DHokerPhong5,
DHokerPhong6}. Here in this subsection we will list some
conventions and all the relevant formulas needed to do explicit
computations for the 4-particle amplitude at two loops.

In the following we will give some formulas in hyperelliptic
representation which will be used later. First all the relevant
correlators are given by\footnote{We follow closely the notation
of \cite{DHokerPhong3}. }
\begin{eqnarray}
\langle \psi^\mu   (z) \psi^\nu   (w) \rangle & = &
-\delta^{\mu\nu} G_{1/2}[\delta] (z,w) = - \delta^{\mu\nu}S_\delta
(z,w),
\nonumber \\
\langle b(z) c(w) \rangle &=& +G_2 (z,w), \nonumber \\
\langle \beta (z) \gamma (w) \rangle &=& -G_{3/2}[\delta] (z,w),
\end{eqnarray}
where
\begin{eqnarray}
& & S_{\delta}(z,w) = { 1\over z-w} \, { u(z) + u(w) \over 2
\sqrt{u(z) u(w) } } , \label{eqszego} \\
& & u(z) = \prod_{i=1}^3 \left( z-A_i \over z-B_i\right)^{1/2}, \\
& & G_2(z,w) = -H(w,z) + \sum_{a=1}^3 H(w,p_a) \, \varpi_a(z,z),
 \label{eq6} \\
& & H(w,z) = { 1\over 2(w- z)} \,\left( 1 + { y(w) \over
y(z) }\right) \, { y(w) \over y(z) }, \\
& & G_{3/2}[\delta](z,w) = - P(w,z) + P(w,q_1) \psi_1^*(z) +
P(w,q_2)\psi_2^*(z), \label{eq51} \\
& & P(w,z) = {1\over \Omega(w)}\, S_{\delta}(w,z)\Omega(z),
\end{eqnarray}
where $\Omega(z)$ is a holomorphic abelian differential satisfying
$\Omega(q_{1,2}) \neq 0$ and otherwise arbitrary. These
correlators were adapted from \cite{Iengo2}. $\varpi_a(z,w)$ are
defined in \cite{DHokerPhong2} and $\psi^*_{1,2}(z)$ are the two
holomorphic $3\over 2$-differentials. When no confusion is
possible, the dependence on the spin structure $[\delta]$ will not
be exhibited. The formulas for the $\langle X(z) X(w) \rangle$ and
related correlators are given in Appendix A.

In order take the limit of $x_{1,2}\to q_{1,2}$ we need the
following expansions:
\begin{eqnarray}
G_{3/2} (x_2, x_1) &=& {1 \over x_1 - q_1} \psi ^* _1 (x_2)
      - \psi ^* _1 (x_2) f_{3/2} ^{(1)} (x_2) +O(x_1 - q_1),
\\
G_{3/2} (x_1, x_2) &=& {1 \over x_2 - q_2} \psi ^* _2 (x_1)
      - \psi ^* _2 (x_1) f_{3/2} ^{(2)} (x_1)  +O(x_2 - q_2),
\end{eqnarray}
for $x_{1,2}  \to q_{1,2}$. By using the explicit expression of
$G_{3/2}$ in (\ref{eq51}) we have
\begin{eqnarray}
f_{3/2} ^{(1)} (q_2) & = & - {\partial_{q_2} S(q_1,q_2) \over
S(q_1,q_2)
} + \partial\psi^*_2(q_2), \label{eq54}\\
f_{3/2} ^{(2)} (q_1) & = &   {\partial_{q_1} S(q_2,q_1) \over
S(q_1,q_2) } + \partial\psi^*_1(q_1) = f_{3/2} ^{(1)}(q_2)|_{ q_1
\leftrightarrow q_2 } . \label{eq55}
\end{eqnarray}

The quantity $\psi^*_\alpha (z)$'s are holomorphic $3\over
2$-differentials and are constructed as follows:
\begin{equation}
\psi^*_\alpha (z) = (z-q_\alpha)S(z,q_\alpha)  \, {y(q_\alpha
)\over y(z)} \, , \qquad \alpha = 1, 2.
\end{equation}
For $z=q_{1,2}$ we have
\begin{eqnarray} &  & \psi^*_\alpha (q_\beta ) =
\delta_{\alpha \beta}, \label{psinormal} \\
& & \partial \psi^*_1 (q_2) = -\partial \psi^*_2 (q_1) =
S(q_1,q_2) = {i\over 4}S_1(q), \\
& & \partial \psi^*_1 (q_1) =  \partial \psi^*_2 (q_2) =
- {1\over 2} \Delta_1(q),  \\
& & \partial^2  \psi^*_1 (q_1) =  \partial^2 \psi^*_2 (q_2) =
{1\over 16}S_1^2(q)  + {1\over 4}\Delta_1^2(q) + {1\over
2}\Delta_2(q),
\end{eqnarray}
where
\begin{eqnarray}
\Delta_n(x) & \equiv & \sum_{i=1}^6 {
1\over (x - a_i)^n }, \\
S_n(x) & \equiv &   \sum_{i=1}^3 \left[ { 1\over (x - A_i)^n } - {
1\over (x - B_i)^n }\right],
\end{eqnarray}
for $  n = 1, 2$. This shows that $\partial\psi^*_\alpha
(q_{\alpha+1})$ and $\partial^2\psi^*_\alpha(q_{\alpha})$ are spin
structure dependent.

For correlators of the $X$ field we refer the reader to
ref.~\cite{AllZhu3} for details.

\subsection{The result of D'Hoker and Phong}

The measure obtained in \cite{DHokerPhong2, DHokerPhong3,
DHokerPhong4, DHokerPhong5} is
\begin{eqnarray}
{\cal A} [\delta] & = & i {\cal Z} \biggl \{ 1  + {\cal X}_1 + {\cal
X}_2 + {\cal X}_3 + {\cal X}_4 +  {\cal X}_5 + {\cal X}_6 \biggr
\},
\nonumber \\
{\cal Z} & = & {\langle  \prod _a b(p_a) \prod _\alpha \delta (\beta
(q_\alpha)) \rangle \over \det \omega _I \omega _J (p_a) } ,
\end{eqnarray}
and the ${\cal X}_i$ are given by:
\begin{eqnarray}
{\cal X}_1 + {\cal X}_6 &=& {\zeta ^1 \zeta ^2 \over 16 \pi ^2}
\biggl [ -\langle \psi(q_1)\cdot \partial X(q_1) \, \psi(q_2)\cdot
\partial X(q_2) \rangle  \nonumber  \\
&& \hskip -1cm
 - \partial_{q_1} G_2 (q_1,q_2) \partial\psi^*_1 (q_2)
 + \partial_{q_2} G_2 (q_2,q_1) \partial\psi^*_2 (q_1)
\nonumber \\
&& \hskip -1cm + 2   G_2 (q_1,q_2) \partial\psi^*_1 (q_2)  f_{3/2}
^{(1)} (q_2) - 2   G_2 (q_2,q_1) \partial\psi^*_2 (q_1)  f_{3/2}
^{(2)} (q_1) \biggr ] \, ,
 \\
{\cal X}_2 + {\cal X}_3 &=&  {\zeta ^1 \zeta ^2 \over 8 \pi ^2}
S_\delta (q_1,q_2) \nonumber \\
&& \hskip  1cm  \times \sum_{a=1}^3 \tilde\varpi_a  (q_1, q_2)
\biggl [ \langle T(\tilde p_a)\rangle + \tilde B_2(\tilde p_a) +
\tilde B_{3/2}(\tilde p_a) \biggr ]\, , \label{eq65}  \\
{\cal X}_4 &=& {\zeta ^1 \zeta ^2 \over 8 \pi ^2} S_\delta
(q_1,q_2) \sum _{a=1}^3 \biggl [ \partial_{p_a} \partial_{q_1} \ln
E(p_a,q_1) \varpi^*_a(q_2) \nonumber \\
& & \hskip  1cm+ \partial_{p_a}
\partial_{q_2} \ln E(p_a,q_2) \varpi ^*_a(q_1) \biggr ]\, ,
 \\
{\cal X}_5 &=& {\zeta ^1 \zeta ^2 \over 16 \pi ^2} \sum_{a=1}^3
\biggl
[ S_\delta (p_a, q_1) \partial_{p_a} S_\delta (p_a,q_2) \nonumber \\
& & \hskip  1cm - S_\delta (p_a, q_2) \partial_{p_a} S_\delta
(p_a,q_1) \biggr ] \varpi_a (q_1,q_2) \, .
\end{eqnarray}

Furthermore, $\tilde B_2$ and $\tilde B_{3/2}$ are given by
\begin{eqnarray}
\tilde B_2(w) & = & -2 \sum _{a=1}^3 \partial_{p_a} \partial_w \ln
E(p_a,w) \varpi^*_a (w) \, , \\
\tilde B_{3/2}(w) &=& \sum_{a=1}^2  G_2 (w,q_a) \partial_{q_a}
\psi^*_\alpha (q_a) + {3 \over 2} \partial_{q_a} G_2 (w,q_a)
\psi^*_\alpha (q_a) \biggr)  \, .
\end{eqnarray}
In comparing with \cite{DHokerPhong4} we have written ${\cal
X}_2$, ${\cal X}_3$ together. We also note that in eq.
(\ref{eq65}) the three arbitrary points $\tilde p_a$ ($a=1,2,3$)
can be different from the three insertion points $p_a$'s of the
$b$ ghost field. The symbol $\tilde\varpi_a$ is obtained from
$\varpi_a$ by changing $p_a$'s to $\tilde p_a$'s. In the next
subsection we will take the limit of $\tilde p_1 \to q_1$. In this
limit we have $\tilde\varpi_{2,3}(q_1,q_2) = 0$ and
$\tilde\varpi_1(q_1,q_2) = -1$. This choice greatly simplifies the
formulas and also make the summation over spin structure doable.

\subsection{The vanishing of the cosmological constant and
non-renormalization theorem}

Before we do any computation of the amplitude we list some useful
formulas:
\begin{eqnarray}
\tilde\varpi_1(q_1,q_2) & = & - { y^2(\tilde p_1)\over y^2(q) }\,
{ (q-\tilde p_2)(q-\tilde p_3) \over  (\tilde p_1-\tilde
p_2)(\tilde p_1-\tilde p_3) } \, ,
\\
\tilde\varpi_2(q_1,q_2) & = & - { y^2(\tilde p_2)\over y^2(q) }\,
{ (q-\tilde p_1)(q-\tilde p_3) \over  (\tilde p_2-\tilde
p_1)(\tilde p_2-\tilde p_3) } \, ,
\\
\tilde\varpi_3(q_1,q_2) & = & - { y^2(\tilde p_3)\over y^2(q) }\,
{ (q-\tilde p_1)(q-\tilde p_2) \over  (\tilde p_3-\tilde
p_1)(\tilde p_3-\tilde p_2) } \, ,
\end{eqnarray}
and
\begin{eqnarray}
\varpi^*_1(u) & = & { y ( p_1)\over y (u) }\, { (u p_1 - {1\over
2}(u+p_1)(p_2+p_3) + p_2p_3)\over (p_1-p_2)(p_1-p_3)} \nonumber
\\
& = & { y ( p_1)\over y (u) } \left[ 1 + { 1\over 2} \, (u-p_1)
\,\left( { 1\over p_1-p_2 }
 + { 1\over p_1-p_3} \right) \right] \, ,
\\
\varpi^*_2(u) & = & { y ( p_2)\over y (u) }\, { (u p_2 - {1\over
2}(u+p_2)(p_3+p_1) + p_1p_3)\over (p_2-p_3)(p_2-p_1)} \, ,
\\
\varpi^*_3(u) & = & { y ( p_3)\over y (u) }\, { (u p_3 - {1\over
2}(u+p_3)(p_1+p_2) + p_1p_2)\over (p_3-p_1)(p_3-p_2)} \, .
\end{eqnarray}
We note here that $\tilde\varpi_1(q_1,q_2) = -1$ and
$\tilde\varpi_{2,3}(q_1,q_2) = 0$ in the limit $\tilde p_1 \to
q_{1,2}$.

The strategy we will follow is  to isolate all the spin structure
dependent parts first. As we will show in the following the spin
structure dependent factors are just $S(q_1,q_2)$,
$\partial_{q_2}S(q_1,q_2)$ and the Szeg\"o kernel if we also
include the vertex operators. Before we do this we will first
write the chiral measure in hyperelliptic language and take the
limit of $\tilde p_1 \to q_1$.

Let's start with ${\cal X}_5$. We have
\begin{equation}
 (S(z,q_1)\partial_zS(z,q_2) -
S(z,q_2)\partial_zS(z,q_1) ) = {i \over 4 (z-q)^2} S_1(z) .
\end{equation}
So the spin structure dependent factor from ${\cal X}_5$ is
effectively $S(z+,z-)$ as shown by the following formulas:
\begin{eqnarray}
S(q_1,q_2) & = & - S(q_2,q_1) = {i \over 4 } \, S_1(q) \, , \\
\partial_{q_2}S(q_1,q_2) & =  &  - \partial_{q_1}S(q_2,q_1)
= - {i\over 8  } S_2(q)  \, .
\end{eqnarray}

For ${\cal X}_4$, the spin structure dependent factor is simply
$S_1(q) \propto S(q_1,q_2)$ as $\ln E(p_a,q_b)$ and
$\varpi^*_a(q_b)$ are spin structure independent and their
explicit expressions   will not be given here.

For ${\cal X}_2 + {\cal X}_3$, we first compute the various
contributions from the different fields. The total stress energy
tensor is:
\begin{eqnarray}
T(z) & = & - {1\over 2}: \partial_zX(z)\cdot \partial_zX(z): +
{1\over 2} :\psi(z)\cdot\partial_z\psi(z): \nonumber \\
& &  - :( \partial b c + 2 b\partial c + {1\over
2}\partial\beta\gamma + {3\over 2}\beta\partial\gamma)(z):
\nonumber \\
& \equiv & T_X(z) + T_{\psi}(z) + T_{bc}(z) + T_{\beta\gamma}(z)
\, ,
\end{eqnarray}
in an obvious notations. The various contributions are
\begin{eqnarray}
T_X(w) & = & -10 T_1(w), \\
T_{\psi}(w) & = & 5 \tilde g_{1/2}(w) =  {5 \over 32}\, (S_1(w))^2, \\
T_{bc}(w) & = & \tilde g_2(w) - 2 \partial_wf_2(w), \\
T_{\beta\gamma}(w) & = & -\tilde g_{3/2}(w) + {3\over 2}\partial_w
f_{3/2}(w),
\end{eqnarray}
where
\begin{eqnarray}
f_2(w) & = & -{3\over 4} \, \Delta_1(w) +
\sum_{a=1}^3 H(w,p_a) \varpi_a(w,w),  \\
\tilde g_2(w) & = & {5 \over 16} \Delta^2_1(w)  + {3\over 8}\,
\Delta_2(w) \nonumber \\
&  & \hskip -1.5cm + \sum_{a=1}^3 H(w,p_a) \varpi_a(w,w) \left(
{ 1\over w-p_{a+1}}+{1\over w-p_{a+2}} - \Delta_1(w) \right) ,\\
f_{3/2}(w) & = & {\Omega'(w)\over \Omega(w)} + {\Omega(q_1)\over
\Omega(w)} \, S(w,q_1)\psi^*_1(w) + {\Omega(q_2)\over \Omega(w)}
\, S(w,q_2)\psi^*_2(w), \label{eq38} \\
\tilde g_{3/2}(w) & = & {1\over 2}\,{\Omega''(w)\over \Omega(w)}
+{1\over 32}\, (S_1(w))^2 \nonumber \\
& & +{\Omega(q_1)\over \Omega(w)} \, S(w,q_1)\partial\psi^*_1(w) +
{\Omega(q_2)\over \Omega(w)} \, S(w,q_2)\partial\psi^*_2(w).
\label{eq39}
\end{eqnarray}

As we said in the last section we will take the limit of $w\to
q_1$. In this limit $T_{\beta\gamma}(w)$ is singular and we have
the following expansion:
\begin{equation}
T_{\beta\gamma}(w)    =   - { 3/2\over (w-q_1)^2} -
{\partial\psi^*_1(q_1)\over w-q_1} -{1\over8}\Delta_1^2(q)  - {
1\over 32}S_1^2(q) + O(w-q_1).
\end{equation}
The dependence on the abelian differential $\Omega(z)$ drops out.
These singular terms are cancelled by similar singular terms in
$\tilde B_{3/2}(w)$. By explicit computation we have: The
dependence on the abelian differential $\Omega(z)$ drops out .
These singular terms are cancelled by similar singular terms in
$\tilde B_{3/2}(w)$. By explicit computation we have:
\begin{eqnarray}
& & \tilde B_{3/2}(w)   =     { 3/2\over (w-q_1)^2} +
{\partial\psi^*_1(q_1)\over w-q_1}   - {1\over 4}\Delta_1^2(q) + {
3\over 4}\Delta_2(q) \nonumber \\
& & \qquad  - \left( {1\over p_1-q} \, { (q-p_2)(q-p_3) \over
(p_1-p_2)(p_1-p_3)} \, \Delta_1(q) + ... \right)
\nonumber \\
& & \qquad - { 3\over 2} \left( {1\over (p_1-q)^2} \,
{(q-p_2)(q-p_3) \over (p_1-p_2)(p_1-p_3)} + ... \right) +
O(w-q_1).
\end{eqnarray}
where $...$ indicates two other terms obtained by cyclic
permutating $(p_1,p_2,p_3)$. By using the above explicit result we
see that the combined contributions of $T_{\beta\gamma}(w)$ and
$\tilde B_{3/2}(w)$ are non-singular in the limit of $w\to q_1$.
We can then take $\tilde p_1\to q_1$ in ${\cal X}_2 + {\cal X}_3$.
In this limit only $a=1$ contributes to ${\cal X}_2+{\cal X}_3$.
This is because $\tilde\varpi_{2,3}(q_1,q_2) = 0$ and
$\tilde\varpi_1(q_1,q_2) = -1$. $T_1(w)$ and $T_{bc}(w)$ are
regular in this limit and spin structure independent. In summary,
the spin structure dependent factors from ${\cal X}_2 + {\cal
X}_3$ are the following two kinds (not including the vertex
operators which will be consider later in section 6):
\begin{equation}
S_1(q)  \propto  S(q_1,q_2), \quad \hbox{and}\quad (S_1(q))^3 .
\end{equation}

Here we note that if we don't take the limit of $w \to q_1$ (or $w
\to q_2$ which has the same effect), the spin  structure dependent
factors from ${\cal X}_2 + {\cal X}_3$ would be much more
complicated. For example we will have a factor of the following
kind:
\begin{equation}
S_1(q) (S_1(w))^2 .
\end{equation}
The summation over spin structure with this factor will give a
non-vanishing contribution and it makes the computation much more
involved. See \cite{AllZhu2} for details.

Finally we come to ${\cal X}_1 + {\cal X}_6$. By using the
explicit results given in eqs. (\ref{eq54})--(\ref{eq55}), we have
\begin{eqnarray} {\cal X}_1 +
{\cal X}_6 & = &    \langle  \partial X(q_1)
 \cdot\partial X(q_2)   \rangle \, S(q_1,q_2)
\nonumber \\
& & - (\partial_{q_1}G_2(q_1,q_2) +
\partial_{q_2}G_2(q_2,q_1) ) S(q_1,q_2)
\nonumber \\
& &   + 2 ( G_2(q_1,q_2) + G_2(q_2,q_1) ) \nonumber \\
& & \times  (\partial\psi^*(q_1) S(q_1,q_2) -
\partial_{q_2}S(q_1,q_2) )  .
\end{eqnarray}
As $G_2(q_1,q_2)$ is spin structure independent, we see that all
the spin structure dependent factors are the following two kinds:
\begin{equation}
 S(q_1,q_2) = {i\over 4} S_1(q),
\end{equation}
and
\begin{equation}
\partial_{q_2} S(q_1,q_2) = {i \over 8} \, S_2(q) .
\end{equation}
Here it is important that the factor $\partial\psi^*_1(q_2)$
cancels the factor $S(q_1,q_2)$ appearing in the denominator of
$f^{(1)}_{3/2}(q_2)$.

From all the above results we see that all the spin structure
dependent parts (for the cosmological constant) are as follows:
\begin{equation}
c_1 S_1(q) + c_2 S_2(q) + c_3 S_1^3(q)+ \sum_{a=1}^3 d_a S_1(p_a).
\label{eqform}
\end{equation}
In computing the $n$-particle amplitude there are more spin
structure factors coming from the correlators of $\psi$. They will
be included at appropriate places.

The vanishing of the cosmological constant is proved by using the
following identities:
\begin{eqnarray}
& & \sum_\delta \eta_\delta Q_\delta S_n(x) = 0, \\
& & \sum_\delta \eta_\delta Q_\delta S_1^3(x) = 0,
\end{eqnarray}
for $n=1,2$ and arbitrary $x$.

For the non-renormalization theorem we need more identities. For
graviton and the antisymmetric tensor the vertex operator is (left
part only):
\begin{equation}
V_i(k_i,\epsilon_i, z_i) = ( \epsilon\cdot \partial X(z_i) + i
k_i\cdot \psi(z_i) \, \epsilon_i\cdot\psi(z_i) )\,  {\rm e}^{ i
k_i \cdot X(z_i, \bar z_i)} .
\end{equation}
Because the vertex operator doesn't contain any ghost fields, all
terms involving ghost fields can be explicit computed which we
have done in the above. For the computation of amplitudes of other
kinds of particles (like fermions), one either resorts to
supersymmetry or can use similar method which was used in
\cite{Sen,Zhu2}.

By including the vertex operators we need to consider the
following extra spin structure dependent terms:
\begin{eqnarray}
\hbox{from}~{{\cal X}_1 + {\cal X}_6}: & & \langle
\psi(q_1)\psi(q_2) \prod_i k_i\cdot \psi(z_i) \,
\epsilon_i\cdot\psi(z_i) \rangle,
\label{eqpsi12} \\
\hbox{from}~{{\cal X}_2 + {\cal X}_3}: & & S_1(q)\, \langle
\psi(q_1) \cdot\psi(q_1) \prod_i k_i\cdot \psi(z_i) \,
\epsilon_i\cdot\psi(z_i) \rangle . \label{eqpsi11}
\end{eqnarray}
The other terms are just the direct product of eq. (\ref{eqform})
with the correlators from the vertex operators $\langle  \prod_i
k_i\cdot \psi(z_i) \, \epsilon\cdot\psi(z_i) \rangle$. Let's study
these direct product terms first.

To prove the non-renormalization theorem we restrict our attention
to 3 or less particle amplitude. For the 3-particle amplitude we
have
\begin{equation}
\langle  \prod_{i=1}^3  k_i\cdot \psi(z_i) \,
\epsilon\cdot\psi(z_i) \rangle \propto
S(z_1,z_2)S(z_2,z_3)S(z_3,z_1) + \hbox{(other terms)}.
\end{equation}
By using the explicit expression of $S(z_1,z_2)$ we have
\begin{eqnarray}
S(z_1,z_2)S(z_2,z_3)S(z_3,z_1) & =  & { 1\over 8
z_{12}z_{23}z_{31}} \left\{ 2 + \left[{u(z_1) \over u(z_2)} +
{u(z_2) \over u(z_1)} \right] \right. \nonumber \\
& & \hskip -1.5cm  \left. + \left[{u(z_1) \over u(z_3)} + {u(z_3)
\over u(z_1)} \right] + \left[{u(z_2) \over u(z_3)} + {u(z_3)
\over u(z_2)} \right] \right\} \, .
\end{eqnarray}
These factors combined with the other factors in eq.
(\ref{eqform}) give vanishing contribution to the $n$-particle
amplitude by using the following  ``vanishing identities":
\begin{eqnarray}
& & \sum_\delta \eta_\delta Q_\delta\left\{ {u(z_1) \over u(z_2)}
+ {u(z_2)\over u(z_1)} \right\}  \, S_n(x)  = 0, \qquad n = 1, 2, \\
& & \sum_\delta \eta_\delta Q_\delta \left\{ {u(z_1) \over u(z_2)}
- (-1)^n {u(z_2)\over u(z_1)} \right\}(S_1(x))^n = 0, \qquad n =
2, 3.
\end{eqnarray}
These identities can be proved by modular invariance and simple
``power counting" which were  explained in detail in
\cite{AllZhu2}.

The terms in eq. (\ref{eqpsi12}) have already been discussed in
\cite{IengoZhu1}. Here we briefly review the argument. We have
\begin{equation}
\langle \psi(q_1)\psi(q_2) \prod_i k_i\cdot \psi(z_i) \,
\epsilon\cdot\psi(z_i) \rangle \propto S(q_1,z_1)
S(z_1,z_2)S(z_2,z_3) S(z_3,q_2)  + \cdots .
\end{equation}
By using the explicit expression of $S(z,w)$ and note that $u(q_2)
= - u(q_1)$ we have
\begin{eqnarray}
& & S(q_1,z_1) S(z_1,z_2)S(z_2,z_3) S(z_3,q_2) \propto
\sum_{i=1}^2 \left[ {u(q_1)\over u(z_i)} - {u(z_i)
\over u(q_1)} \right]  \nonumber \\
& & \qquad + {u(z_1)\over u(z_2)} - {u(z_2) \over u(z_1)} +
{u(z_1)\over u(z_3)} - {u(z_3) \over u(z_1)}  \nonumber \\
& & \qquad + {u(z_2)\over u(z_3)}- {u(z_3) \over u(z_2)} + {u(q_1)
u(z_2)\over u(z_1) u(z_3)}
 - {u(z_1) u(z_3) \over u(q_1) u(z_2)} .
\end{eqnarray}
These terms also give  vanishing  contributions as we can prove
the following identities:
\begin{eqnarray}
& & \sum_\delta \eta_\delta Q_\delta\left\{ {u(z_1) \over u(z_2)}
- {u(z_2)\over u(z_1)} \right\}     = 0,   \label{eq70} \\
& & \sum_\delta \eta_\delta Q_\delta \left\{ {u(z_1) u(z_2) \over
u(z_3) u(z_4)} -   {u(z_3) u(z_4)\over u(z_1) u(z_2)} \right\}  =
0.
\end{eqnarray}
These identities were firstly proved in \cite{IengoZhu1}. The
proof is quite simple by using modular invariance. For example we
have
\begin{eqnarray}
& & \sum_\delta \eta_\delta Q_\delta\left\{ {u(z_1) \over u(z_2)}
- {u(z_2)\over u(z_1)} \right\}
\nonumber \\
& & = { 1\over y(z_1) y(z_2) } \sum_\delta \eta_\delta
Q_\delta\left\{ \prod_{i=1}^3(z_1-A_i)(z_2-B_i) -
\prod_{i=1}^3(z_1-B_i)(z_2-A_i) \right\}
\nonumber \\
& & \propto {(z_1 - z_2)\, P(a) \over y(z_1)y(z_2)} ,
\end{eqnarray}
which must be vanishing as the degrees of the  homogeneous
polynomials (in $a_i$ and $z_j$) don't match. Here we have  used
again the modular invariance of the above expression.\footnote{The
minus sign in eq. (\ref{eq70}) makes the expression invariant
under the all the modular transformations. With a plus sign the
expression is only invariant under a subgroup of the modular
transformation. Nevertheless eq. (\ref{eq70}) is still true with a
plus sign. The explicit results are:
\begin{eqnarray}
& & \sum_\delta \eta_\delta Q_\delta\left\{ {u(z_1) \over u(z_2)}
+ {u(z_2)\over u(z_1)} \right\}     = 0,   \\
& & \sum_\delta \eta_\delta Q_\delta \left\{ {u(z_1) u(z_2) \over
u(z_3) u(z_4)} + {u(z_3) u(z_4)\over u(z_1) u(z_2)} \right\}  = {
2 P(a) z_{13}z_{14}z_{23}z_{24}\prod_{i=1}^4(a_1 -z_i) \over
 \prod_{i=1}^4y(z_i) \prod_{i=2}^6(a_1-a_i)} .
\end{eqnarray} }

The last term we need to compute is the term in eq.
(\ref{eqpsi11}). We have
\begin{eqnarray}
& & \langle :\psi(q_1) \cdot \partial \psi(q_1): \prod_i k_i\cdot
\psi(z_i) \, \epsilon\cdot\psi(z_i) \rangle_c
   =  K(1,2,3) \nonumber \\
& &  \qquad  \times (S(x,z_1,z_2,z_3) + S(x,z_2,z_3,z_1) +
S(x,z_3,z_1,z_2)
 \nonumber \\
& &  \qquad \quad - S(x,z_1,z_3,z_2) - S(x,z_2,z_1,z_3) -
S(x,z_3,z_2,z_1) ), \label{eqpsi33}
\end{eqnarray}
where
\begin{eqnarray}
K(1,2,3) & = & k_1\cdot\epsilon_3 k_2\cdot\epsilon_1
k_3\cdot\epsilon_2- k_1\cdot\epsilon_2 k_2\cdot\epsilon_3
k_3\cdot\epsilon_1
\nonumber \\
& & + k_1\cdot k_2(k_3\cdot\epsilon_1\epsilon_2\cdot\epsilon_3 -
k_3\cdot\epsilon_2\epsilon_1\cdot\epsilon_3) \nonumber \\
& & + k_2\cdot k_3(k_1\cdot\epsilon_2\epsilon_3\cdot\epsilon_1-
k_1\cdot\epsilon_3\epsilon_2\cdot\epsilon_1) \nonumber \\
& & + k_3\cdot k_1(k_2\cdot\epsilon_3\epsilon_1\cdot\epsilon_2-
k_2\cdot\epsilon_1\epsilon_3\cdot\epsilon_2)    .
\end{eqnarray}
We note that $K(1,2,3)$ is invariant under the cyclic permutations
of (1,2,3). It is antisymmetric under the interchange $2
\leftrightarrow 3$. We  have used these properties in eq.
(\ref{eqpsi33}).

To compute explicitly these expressions we first note the
following:
\begin{equation}
\partial_x S(z, x) = { 1\over  2(z-x)^2 }
\, {u(z) + u(x) \over \sqrt{u(z)u(x)}  }  - {S_1(x) \over 8\,
(z-x) } \, {u(z) - u(x) \over \sqrt{u(z)u(x)} } \,   .
\label{eqpartial}
\end{equation}
In order to do the summation over spin structure  we need a
``non-vanishing identity". This and other identities needed in the
4-particle amplitude computations are summarized as follows:
\begin{eqnarray}
& &  \sum_\delta \eta_\delta Q_\delta \left\{ {u(z_1)  u(z_2)
\over u(z_3)u(z_4)} - (-1)^n {u(z_1)u(z_2)\over u(z_3)u(z_4) }
\right\} (S_m(x))^n \nonumber  \\
& &  \qquad \qquad =  {2 P(a) \prod_{i=1}^2\prod_{j=3}^4 (z_i-z_j)
\prod_{i=1}^4(x-z_i) \over y^2(x) \prod_{i=1}^4 y(z_i) } \times
C_{n,m},
\end{eqnarray}
where
\begin{eqnarray}
C_{1,1} & = & 1,  \label{eq991} \\
C_{2,1} & = & - 2 (\tilde z_1 + \tilde z_2 - \tilde z_3 - \tilde
z_4) ,     \\
C_{1,2} & = & \Delta_1(x)  - \sum_{k=1}^4 \tilde z_k   , \\
C_{3,1} & = & 2 \Delta_2(x) - \Delta_1^2(x) + 2 \Delta_1(x)\,
\sum_{k=1}^4 \tilde z_k \nonumber \\
& &  + 4 \left(\tilde z_1 \tilde z_2   - 2 ( \tilde z_1  + \tilde
z_2 )(\tilde z_3 + \tilde z_4 ) +
\tilde z_3  \tilde z_4  \right) \nonumber \\
&   \equiv  & C_{3,1}(z_1,z_2,z_3,z_4,x), \label{eq992} \\
\tilde z_k & = & { 1\over x - z_k}, \\
 P(a) & = & \sum_{i<j}(a_i-a_j).
\end{eqnarray}
$C_{1,1}$ and $C_{1,2}$ were derived in \cite{IengoZhu2}. Although
other values of $n,m$ also gives modular invariant expressions,
the results are quite complex.\footnote{This is due to the
non-vanishing of the summation over spin structures when we set
$z_1=z_3$ or $z_1=z_4$, etc.} Fortunately we only need to use the
above listed results.

By using these formulas we have:
\begin{equation}
\sum_\delta \eta_\delta Q_\delta S(x,z_1,z_2,z_3) \,  S_1(x)  = -
{P(a)\over16  y^2(x) } \prod_{i=1}^3 {x-z_i\over y(z_i)} .
\end{equation}
We note that the above formula is invariant under the interchange
$z_i \leftrightarrow z_j$.

By using this result and   eq. (\ref{eqpsi33}), we have:
\begin{equation}
\sum_\delta \eta_\delta Q_\delta S(q_1,q_2) \langle
\psi(q_1)\cdot\partial\psi(q_1)\prod_{i=1}^3 k_i\cdot\psi(z_i)
\epsilon_i\cdot\psi(z_i)\rangle_\delta =0.
\end{equation}
This completes our verification of the non-renormalization theorem
at two loops.

\subsection{The 4-particle amplitude}

To explicitly compute the 4-particle amplitude we need the
following summation formulas:
\begin{eqnarray}
& &  \sum_{\delta}\eta_\delta Q_\delta  \, S_n(x)
S(z_1,z_2,z_3,z_4) = \sum_{\delta}\eta_\delta Q_\delta  \, S_n(x)
(S(z_1,z_2)S(z_3,z_4))^2 \nonumber \\
&   & \qquad = { P(a) \over y^2(x) }\, \prod_{i=1}^4 { x-z_i \over
y(z_i)} \, \left\{
\begin{array}{ll}
1,  & n =1, \\
\Delta_1(x) - \sum_{i=1}^4 \tilde z_i, & n=2.
\end{array}
\right. \label{eqold00} \\
& & \sum_\delta \eta_\delta Q_\delta (
S(x,z_1)S(z_1,z_2)\partial_xS(z_2,x) + (z_1\leftrightarrow z_2))
\, (S(z_3,z_4))^2 \, S_1(x)   \nonumber \\
& & \qquad = - {P(a)\over 16\,  y^2(x) }\, \prod_{i=1}^4{
x-z_i\over  y(z_i)} \, \left( \tilde z_{14}\tilde z_{23} + \tilde
z_{13}\tilde z_{24} \right) ,
\\
& & \sum_\delta \eta_\delta Q_\delta (
S(x,z_1)S(z_1,z_2)S(z_2,z_3)S(z_3,z_4)\partial_xS(z_4,x) \nonumber
\\
& &  \qquad \quad + (z_1\leftrightarrow z_4, z_2\leftrightarrow
z_3)) \, \, S_1(x)   =  {P(a)\over 16\,  y^2(x)}\, \prod_{i=1}^4{
x-z_i\over  y(z_i)} \,  \tilde z_{14} \tilde z_{23} , \\
& & \sum_\delta \eta_\delta Q_\delta  (S(z_1,z_2)S(z_3,z_4))^2
(  S_1(x)  )^3 \nonumber \\
& & \qquad =  {P(a)  \over 8\, y^2(x)}\, \prod_{i=1}^4{ x-z_i\over
y(z_i)} \, \left[ 2 \Delta_2(x) - \Delta_1^2(x) +
2 \Delta_1(x)\sum_{i=1}^4{\tilde z_i}  \right. \nonumber \\
& & \qquad \qquad \left. + 4 \left(\sum_{k<l} {\tilde z_k\tilde
z_l } - {6\tilde z_1\tilde z_2 } - {6\tilde z_3\tilde z_4 }
\right)\right],
\nonumber \\
& & \sum_\delta \eta_\delta Q_\delta
S(z_1,z_2)S(z_2,z_3)S(z_3,z_4)S(z_4,z_1) ( S_n(x)  )^3 \nonumber \\
& & \qquad    =  {P(a) \over 8 \, y^2(x)}\, \prod_{i=1}^4{
x-z_i\over  y(z_i)} \, C_{3,1}(z_1,z_3,z_2,z_4,x) .
\end{eqnarray}
The first formula was derived in \cite{IengoZhu2, Zhu}. All the
rest formulas can be derived by using the formulas given in eqs.
(\ref{eq991})--(\ref{eq992}) and eq. (\ref{eqpartial}) for the
derivative of the Szeg\"o kernel.

After a long calculation we have the following expression for the
chiral integrand \cite{AllZhu1, AllZhu3}:
\begin{eqnarray}
{\cal A} & = & c_{II} \, K(k_i,\epsilon_i) \langle :( \partial
X(q_1) + X(q_2) ) \cdot ( \partial X(q_1) +
\partial X(q_2) ) : \nonumber \\
& &  \times \prod_{i=1}^4 \hbox{e}^{i k_i \cdot X(z_i, \bar z_i)}
\rangle \prod_{i=1}^4 { q -z_i\over y(z_i) }  \nonumber \\
& = &  { c_{II} \, K(k_i,\epsilon_i) \over   \prod_{i=1}^4 y(z_i)
} \,
\prod_{i<j} |E(z_i,z_j)|^{2   k_i\cdot k_j} \nonumber \\
&  & \times ( s (z_1z_2 + z_3 z_4) + t(z_1z_4+ z_2 z_3) + u(z_1z_3
+ z_2 z_4)) , \label{eq777}
\end{eqnarray}
where $ K(k_i,\epsilon_i)$ is the standard kinematic factor
appearing at tree level and one loop computations
\cite{GreenSchwarz1, Zhu, IengoZhu2}. Here we used the explicit
correlators for $\langle \partial X(z) \partial X(w)$ and $\langle
\partial X(z) X(w,\bar w)$ given in \cite{Knizhnik, Zhu}. As
 it is expected, the find result is
independent on the insertion points $q_{1,2}$ and $p_a$'s.

For type II superstring theory one can combine the left part and
the right part. The final result would be as follows:
\begin{eqnarray}
{\cal A}_4 & = &   K(k_i,\epsilon_i) \langle :(
\partial X(q_1) + \partial X(q_2) ) \cdot ( \partial X(q_1) +
\partial X(q_2) )   \nonumber \\
& &  \times \, ( \bar \partial X(\bar p_1) + \partial X(\bar p_2)
) \cdot ( \bar \partial X(\bar p_1) + \bar \partial X(\bar p_2) )
: \nonumber \\
& & \times  \prod_{i=1}^4 \hbox{e}^{i k_i \cdot X(z_i, \bar z_i)}
\rangle \prod_{i=1}^4 { (q -z_i)(\bar p - \bar z_i)
\over |y(z_i)|^2 }  \nonumber \\
& = &   { K(k_i,\epsilon_i) \over   \prod_{i=1}^4 |y(z_i)|^2  } \,
\prod_{i<j} |E(z_i,z_j)|^{2   k_i\cdot k_j} \nonumber \\
&  & \times | s (z_1z_2 + z_3 z_4) + t(z_1z_4+ z_2 z_3) + u(z_1z_3
+ z_2 z_4)|^2  . \label{eq7777}
\end{eqnarray}

The amplitude is obtained by integrating over the moduli space. At
two loops, the moduli space can be parametrized either by the
period matrix or three of the six branch points. We have
\begin{eqnarray}
{A}_{II}(k_i,\epsilon_i) & = &  c_{II}\, K(k_i,\epsilon_i) \, \int
{ \prod_{i=1}^6 {\rm d}^2 a_i/{\rm d} V_{pr} \over T^5  \,
\prod_{i<j} |a_i - a_j|^2 } \,  \nonumber \\
& & \times \prod_{i=1}^4 {  { \rm d}^2 z_i \over |y(z_i)|^2 } \,
\prod_{i<j} {\rm exp}\left[ - k_i\cdot k_j \ G(z_i,z_j) \right]
\nonumber \\
&  & \hskip -1cm \times | s (z_1z_2 + z_3 z_4) + t(z_1z_4+ z_2
z_3) + u(z_1z_3 + z_2 z_4)|^2 , \label{eqr95}
\end{eqnarray}
where  $c_{II}$ is a constant which should be determined by
factorization or unitarity (of the $S$-matrix). The amplitude is
an integration of the above integrand over the moduli space.

An immediate application of the above result is to study the
perturbative correction to the $R^4$ term at two loops. In the low
energy limit $k_i \to 0$, the chiral integrand is also 0 apart
from the kinematic factor because of some extra factors of $s$,
$t$ and $u$ in eq. (\ref{eq7777}). This confirms the explicit
computation of Iengo \cite{Iengo} by using the old result
\cite{IengoZhu2, Zhu}, and it is in agreement with the indirect
argument of Green and Gutperle \cite{GreenGutperle}, Green,
Gutperle and Vanhove \cite{Green2}, and Green and Sethi
\cite{GreenSethi}. Our new result also explicitly verifies the
claim given in the the Appendix B of \cite{Stieberger}.

\section*{Acknowledgments}

Chuan-Jie Zhu would like to thank Roberto Iengo, Zhu-Jun Zheng and
Jun-Bao Wu for fruitful collaborations.   He would also like to
thank E. D'Hoker and D. Phong for discussions and Jian-Xin Lu and
the hospitality at the Interdisciplinary Center for Theoretical
Study, Physics, University of Science and Technology of China.

\end{document}